# Significant tuning of dispersive mode coupling in doubly clamped MEMS beam resonators using thermally-induced buckling effect

Chao Li[1], Qian Liu[1], Kohei Uchida[1], Hua Li[2], Kazuhiko Hirakawa[3], and Ya Zhang[1,a)]

[1]*Institute of Engineering, Tokyo University of Agriculture and Technology, Koganei-shi, Tokyo, 184-8588, Japan*

[2]*Shanghai Institute of Microsystem and Information Technology, Chinese Academy of Sciences, Shanghai 200050, China*

[3]*Institute of Industrial Science, University of Tokyo, 4-6-1 Komaba, Meguro-ku, Tokyo 153-8505, Japan*

## Abstract

Dispersive mode coupling is a promising mechanism for the development of advanced micro/nanoelectromechanical devices. However, strong coupling strength remains a key challenge limiting the practical applications of dispersive mode coupling effect. Here, we experimentally demonstrate the significant tuning of the mode coupling coefficient of two flexural vibrational modes in a doubly-clamped MEMS beam resonator using thermally-induced buckling effect, which enables variable coupling strengths to be implemented for practical applications. Furthermore, a theoretical model is developed to describe the mode coupling coefficient, showing that the tunability is owing to the breakdown of the symmetric shape of the MEMS beam caused by buckling. Moreover, the theoretical model defines a simple relation between the coupling coefficient and the nonlinearity of the two coupled modes. These results provide valuable insight into physical mechanisms underlying dispersive mode coupling effect, as well as pave the way for the development of advanced MEMS devices with application-specified mode coupling strength.

---

a) Electronic mail: zhangya@go.tuat.ac.jp

# Introduction

Interactions among two or more vibrational modes in micro/nanoelectromechanical system (M/NEMS) resonators have garnered significant research interest over the past decade. Two primary types of intermodal interactions in MEMS resonators have been intensively investigated [1]. One is known as the internal resonance, where two modes are coupled when their resonance frequencies fulfil an integer ratio [2–4], allowing the coherent energy exchange between the two modes. This phenomenon has been used for frequency locking [5–7], synchronized oscillation [8–10], signal amplification [11,12], energy dissipation [13–15], and highly-sensitive sensing [2,16–18]. The second type of interaction is referred as dispersive mode coupling, which involves the change in the resonance frequency of one mode by the oscillation of another mode [19,20], through the vibration-induced tension change in MEMS beam. Unlike the internal resonance, the dispersive mode coupling is not restricted to any specific frequency conditions, allowing it to occur between any two modes. This flexibility provides a significant advantage in bridging distinct mechanical modes for various applications. However, the energy exchange in dispersive mode coupling is generally more modest compared to internal resonance, making it difficult to achieve strong mode coupling. Tuning of mode coupling strength (i.e., mode coupling coefficient) in M/NEMS resonators remains a key challenge limiting practical applications of dispersive mode coupling effect.

Although mode coupling mechanisms based on the geometric nonlinearity or the vibration-induced tension in M/NEMS resonators have been reported in the past studies [19–23], these studies so far have mostly concentrated on experimental observations. Investigations into the quantitative tuning of dispersive mode coupling strength are less touched, and the relation between coupling and nonlinearity is also unclear. Furthermore, manufacturing defects and small asymmetries can have a huge impact in such systems. Consequently, the ability to determine coupling strength by geometric design remains limited and is often elusive.

Here, we demonstrate the significant tuning of the mode coupling strength in a doubly-clamped MEMS resonator using thermally-induced buckling effect. Two flexural vibrational modes are mechanically coupled through the additional tension induced by the vibrations of the MEMS beam, and the tuning of mode coupling strength is achieved experimentally by inducing the buckling of the beam with thermal effect. The buckling breaks the symmetry of the MEMS beam, which changes the vibration-induced additional tension of the MEMS beam, thus gives the tunability in the coupling strength between the two modes. Moreover, we present a general theoretical model based on Euler-Bernoulli beam theory to quantitatively describe the mode coupling coefficient, which shows nice agreement with the experimental result. Compared to other buckling-based models [24,25], we focus on how the buckling effect modulates the dispersive mode coupling and clarify the mechanism for controlling the coupling coefficient. Furthermore, the model gives a simple relation between the mode coupling coefficient and the mechanical nonlinearity of the two vibrational modes, which enables the prediction of the coupling coefficient from simple Duffing nonlinearity measurements. These results provide valuable insight for understanding the physical origins of the mode coupling effect in MEMS resonators, as well as pave the way for achieving strong or weak coupling implemented in practical device applications.

## Results

**Vibration modes for the mode coupling measurement**

The measurements were performed on a GaAs doubly clamped MEMS beam resonator [26–28] fabricated by using a modulation-doped (Al, Ga)As/GaAs heterojunction structure grown by molecular-beam epitaxy [29], as schematically shown in Fig. 1(a). The MEMS beam consists of a GaAs/$Al_{0.3}Ga_{0.7}$As superlattice buffer layer, a 1-μm-thick GaAs layer, and a 2-dimensional electron gas (2DEG) layer. Beneath the beam, there is a 3-μm-thick $Al_{0.7}Ga_{0.3}$As sacrificial layer on a (100)- oriented GaAs substrate. The suspended beam structure as shown in the upper inset of Fig. 1(a) is formed by selectively etching the sacrificial layer with diluted hydrofluoric acid (HF). Figure 1(b) shows a microscope image of the device, which has a geometry of 133 μm($L$)×27 μm($b$)×1 μm($h$). Top metal gates ($C_1$ and $C_2$) are formed on the beam to drive mechanical vibrations using the piezoelectric effect of GaAs. A 15-nm-thick NiCr layer was deposited on the beam as a heater for generating thermal strain in the MEMS beam.

The measurement system consists of a laser Doppler vibrometer (LDV) and a lock-in amplifier with a built-in phase locked loop (PLL). We drive the beam into vibration by applying an ac voltage ($V_D$) to one of the piezoelectric capacitors ($C_1$ or $C_2$) and then measure the beam vibration by the LDV and the lock-in amplifier. Figure 1(c) plots the measured oscillation spectrum with $V_D$ = 100 mV. As seen, the first three vibrational modes obtained are 1st bending mode (235.5 kHz), 2nd bending mode (644.5 kHz) and 1st torsional mode (752 kHz), whose mode shapes are schematically shown by the inset of Fig. 1(c). In this work, we utilize the first two flexural modes (i.e., 1st and 2nd bending modes) to study their mode coupling, since they are more sensitive to the thermal strain along the beam axis [19,23] than the torsional modes.

We applied a DC voltage ($V_{NiCr}$) to the NiCr heater (see Fig. 1(b)) to induce a heat in the beam, which reduces the resonance frequency through the thermally induced axis strain [30,31]. Figures 1(d) and (e) show the measured resonance frequencies of the first two modes as a function of heating power, $P$. As seen, for the 1st bending mode (Fig. 1(d)), the resonance frequency first decreases with the $P$ (pre-buckling: $P$ <1.33 mW), then levels off and starts to increase with the further increased $P$, suggesting that the beam reaches its onset of critical buckling at $P$ =1.33 mW and subsequently enters its post-buckling regime. For an ideally straight beam, the bending moment and the flexural rigidity cancel each other out at the critical buckling point, giving a quasi-zero resonance frequency for the 1st bending mode. In experiments, however, since there is always a small initial center deflection, $x_0$, due to the mesa structure of the MEMS beam, the frequency does not drop to zero [26,30]. From the frequency shifts shown in Figs. 2(d) and (e), we can calibrate that $x_0 =\sim 100$ nm for the present MEMS beam, which plays an important role in building an accurate mode coupling model. For the 2$^{nd}$ bending mode (Fig.1(e)), the frequency keeps decreasing with $P$, showing that this mode is less affected by the beam buckling than the 1$^{st}$ bending mode.

### Experimental measurement of mode coupling coefficient

We follow the method reported in Refs [19,23] to measure the mode coupling coefficient. The 1$^{st}$ bending mode (mode-1) is used as a probe mode, which is driven into a self-sustained oscillation by the PLL [32]. The 2$^{nd}$ bending mode (mode-2) is employed as the pump mode. When mode-2 is driven into oscillation, it gives an additional tension to the MEMS beam, thus induces a shift ($\Delta f$) in the resonance frequency of mode-1. Such coupling behavior can be formulated by

$$\Delta f = f_1' - f_1 = f_1 \lambda_{1,2} a_2^2, \qquad (1)$$

where $f_1$, $f_1'$ are the resonance frequencies of mode-1 without and with the excitation of mode-2, respectively, $a_2$ is the oscillation amplitude of mode-2, and $\lambda_{1,2}$ is the mode coupling coefficient between the two modes. Here, a small driving voltage ($V_D$=60 mV) is applied to mode-1 to keep its oscillation amplitude in the linear regime (typically, ~18 nm). Furthermore, during the pump measurement, the oscillation amplitude of mode-1 remains stable with fluctuations below 1 nm. This ensures that the observed $\Delta f$ is primarily due to the coupling effect with mode-2, rather than the intra-mode nonlinearities of mode-1.

By applying various heating powers, $P$, and sweeping the driving frequency near the resonance frequency of mode-2, we simultaneously measure the $a_2$ and the $f_1'$. Figures 2(a-g) plot the measured $a_2$ and Figs. 2(h-n) plot the measured $f_1'$, as a function of the driving frequency, respectively. It can be clearly seen that, when mode-2 is driven into oscillation, there is a frequency shift in the resonance frequency of mode-1, indicating that the dispersive mode coupling occurs between the two modes. Furthermore, when $P \leq 1.32$ mW as shown in Figs. 2(a-c) and (h-j), $f_1'$ shows a blue shift as $a_2$ increases, indicating that the $\lambda_{1,2} >0$ under this condition. However, when $P =1.35$ mW as shown in Figs. 2(d) and (j), the $f_1'$ exhibits a tiny back-and-forth fluctuation as mode-2 is driven into oscillation, indicating that mode-2 barely affects the resonance frequency of mode-1, and the coupling becomes rather tiny under this condition ($\lambda_{1,2} =\sim 0$). Moreover, when $P \geq 1.38$ mW as shown in Figs. 2(e-g) and (l-n), the $f_1'$ presents a red shift as $a_2$ increases, suggesting that $\lambda_{1,2} <0$ under this condition. The above results demonstrate that the mode coupling coefficient between the pump and probe modes have been largely tuned by the input heat to the MEMS beam.

To further quantitatively characterize the tunability of the $\lambda_{1,2}$, we calculate $\lambda_{1,2}$ by performing linear fitting with Eq. (1); the data is from the measured $a_2$ and $f_1'$, at various $P$, as exemplified by Fig. 2. Figures 3 (a-c) show the representative fitting results for the cases of $\lambda_{1,2} >0$, $\lambda_{1,2} \approx 0$ and $\lambda_{1,2} <0$, respectively. As seen, all of them show good linear relationship between $f_1'$ and $a_2^2$, which well agrees with Eq. (1). The measured $\lambda_{1,2}$ at $P = 0$ is $1.53 \times 10^{12}$ m$^{-2}$, giving a quantitative relation between the oscillation amplitude of mode-2 and the resonance frequency shift of mode-1. On the other hand, the measured $\lambda_{1,2}$ in Fig. 3(b) is only $7.91 \times 10^{10}$ m$^{-2}$, which is less than 10% of the measured $\lambda_{1,2}$ in Figs. 3(a) and (c). Thus, it can be considered as a quasi-zero coupling coefficient. The measured $\lambda_{1,2}$ at various $P$ are shown as the black solid dots in Fig. 3(d). As seen, the $\lambda_{1,2}$ has been significantly tuned from positive to negative values, demonstrating our method a feasible approach to tune the mode coupling coefficient. It is worth noting that there exists a point that $\lambda_{1,2} = 0$, where the two modes decoupled from each other, showing the potential in the applications that require two modes vibrating independently.

Furthermore, we observed an interesting sudden drop in $\lambda_{1,2}$ at $P = 0.78$ mW as indicated by black arrow in Fig. 3(d). A blowup of the corresponding area (0.6mW < P < 0.9 mW) is shown in Fig. 4(a) for clarification. At the dropping point for $\lambda_{1,2}$, the resonance frequencies of the two modes roughly fulfill an integer ratio $f_1$: $f_2$=1:3 ($f_1$=~190 kHz and $f_2$=~570 kHz), triggering the internal resonance between the two modes. Thus, the sudden drop may be owing to 1:3 internal mode resonance [1,28,33,34], formed by the 3$^{rd}$ order nonlinearity of the lower frequency mode (mode-1).

To verify the relation between the drop in $\lambda_{1,2}$ and the 1:3 internal resonance, we plot the peak oscillation amplitude of mode-2 ($a_{2max}$) as a function of its resonance frequency in the thermal tuning measurement, as shown in Fig. 4(b). As seen, $a_{2max}$ shows a gradual decrease when the resonance frequency is thermally-modulated to the low frequency side, which is owing to the reduction of the Q-factor in the thermal tuning process [26]. However, $a_{2max}$ shows a notable increase when $f_2$ is modulated to ~570 kHz, suggesting the presence of 1:3 internal resonance at this frequency. We also performed an open-loop measurement for mode-1 to support our analysis. We apply $P = 0.8$ mW to the beam and strongly drive mode-1 into the nonlinear regime, as internal mode resonance is stronger at larger oscillation amplitudes. Figure 4 (c) shows the measured oscillation amplitude (black) and phase (red) of the mode-1 versus the driving frequency. Clear drops in the oscillation amplitude of mode-1 are observed at ~190 kHz, which coincides with the frequency where the measured $\lambda_{1,2}$ shows notable drop, suggesting their correlation. However, since mode-1 was driven under a much weaker condition in the $\lambda_{1,2}$ measurements, it remains unclear how the internal mode resonance influences dispersive mode coupling under the current experimental conditions.

**Theoretical model of dispersive mode coupling**

To understand the physical origin of the change in $\lambda_{1,2}$, we developed a theoretical model for the doubly-clamped MEMS beam resonator as schematically shown in Fig. 5(a). The model consists of a doubly-clamped beam with an initial transverse displacement, $X_0(u)$, governed by [35]

$$X_0(u) = x_0\phi_0(u) = -\frac{x_0}{2}\left(1 - \cos\frac{2\pi}{L}u\right), \tag{2}$$

where $x_0$ is the initial center deflection of the beam; $\phi_0(u)$ is the profile function of initial shape; $u$ is the coordinate along the length of the beam. Note that when a thermal strain is applied, the center deflection will be increased, which is expressed by $x_T$. We assume there are two modes (mode-2 for the pump mode and mode-1 for the probe mode) excited on the beam, and their mode shape functions are as [36],

$$\phi_1 = 1.01781 \cos[4.73004u] - 1.01781 \cosh[4.73004u] \\ - \sin[4.73004u] + \sinh[4.73004u] \tag{3}$$

$$\phi_2 = 0.999223 \cos[7.8532u] - 0.999223 \cosh[7.8532u] \\ - \sin[7.8532u] + \sinh[7.8532u], \tag{4}$$

The calculated 1-dimensional mode shapes using Eqs. (3) and (4) are shown in the upper insets of Fig. 5(a). The effect of mode-2 on the dynamical behavior of mode-1 can be described by the following Duffing motion equation, as (more details is presented in Supplementary Note 1 [37])

$$\ddot{x}_1 + (\omega_1^2 + D_1)x_1 + \beta_1 x_1^2 + \alpha_1 x_1^3 + D_2 = 0 \tag{5}$$

where $\omega_1$ is the resonance frequency of mode-1, $\beta_1 = \frac{3x_T E}{2\rho L^4} W_{1,1} W_{0,1}$ and $\alpha_1 = \frac{E}{2\rho L^4} W_{1,1}^2$ are the quadratic and cubic nonlinearity coefficients of mode-1, respectively. In the same spirit, $\alpha_2 = \frac{E}{2\rho L^4} W_{2,2}^2$ is for the cubic (hardening) nonlinearity coefficient of mode-2. $D_1 = \frac{EW_{1,1}W_{2,2}}{2\rho L^4}\frac{a_2^2}{2}$ and $D_2 = \frac{x_T E}{2\rho L^4} W_{2,2} W_{0,1} \frac{a_2^2}{2}$ are the terms with respect to the oscillation amplitude of mode-2 ($a_2$) that affect the dynamical behavior of mode-1. $E$ is the Young's modulus, $\rho$ and $L$ are the density and beam length, respectively. $W_{2,2}$, $W_{1,1}$ and $W_{0,1}$ are the mode overlapping parameters determined by the eigen vibrational mode shapes and the initial shape of the beam, as,

$$W_{mn} = \int_0^1 \frac{\partial \phi_m}{\partial u}\frac{\partial \phi_n}{\partial u} du \quad (m, n = 0,1,2), \tag{6}$$

Next, we take a solution of the form for Eq. (5), as

$$x_1 = \frac{1}{2}\left(a_1 e^{i\omega_1't} + \overline{a_1} e^{-i\omega_1't}\right) + \delta, \tag{7}$$

where $a_1$ indicates the complex oscillation amplitude of mode-1, $\omega_1'$ indicates the oscillation frequency of mode-1 under a certain driving force, and $\delta$ is a perturbed parameter given by $D_1$ and $D_2$. By substituting Eq. (7) into Eq. (5), dropping terms in higher order of $\delta$, and then matching the terms in $e^{i\omega_1't}$ and constant terms, respectively, this yields the following two equations,

$$-\omega_1'^2 + \omega_1^2 + D_1 + \frac{3}{4}\alpha_1 a_1^2 + 2\beta_1 \delta = 0, \tag{8}$$

$$\delta\omega_1^2 + \delta D_1 + \frac{3}{2}\delta\alpha_1 a_1^2 + \frac{1}{2}\beta_1 a_1^2 + D_2 = 0. \tag{9}$$

From Eq. (9) we can obtain:

$$\delta = \frac{-\frac{1}{2}\beta_1 a_1^2 - D_2}{\omega_1^2 + D_1 + \frac{3}{2}\alpha_1 a_1^2}, \tag{10}$$

in which $D_1 + \frac{3}{2}\alpha_1 a_1^2$ can be can be ignored since it is a much smaller value compared to $\omega_1^2$. Thus,

$$\delta \approx \frac{-\frac{1}{2}\beta_1 a_1^2 - D_2}{\omega_1^2}. \tag{11}$$

Then by substituting Eq. (11) into Eq. (8), it yields

$$\frac{\omega_1'}{\omega_1} = \sqrt{1 + \frac{D_1}{\omega_1^2} + \frac{3\alpha_1}{4\omega_1^2}a_1^2 - \frac{\beta_1^2 a_1^2}{\omega_1^4} - \frac{2\beta_1 D_2}{\omega_1^4}}, \tag{12}$$

Taking Taylor expansion to $\omega_1'$, we can obtain:

$$\omega_1' \approx \omega_1 \left[1 + \left(\frac{3\alpha_1}{8\omega_1^2} - \frac{\beta_1^2}{2\omega_1^4}\right)a_1^2 + \left(\frac{EW_{1,1}W_{2,2}}{8\rho L^4 \omega_1^2} - \frac{\beta_1 x_T EW_{2,2}W_{0,1}}{4\rho L^4 \omega_1^4}\right)a_2^2\right]. \tag{13}$$

This equation shows the frequency shift of mode-1 ($\Delta\omega = \omega_1' - \omega_1$) as a function of the oscillation amplitudes of the two modes ($a_1$ and $a_2$), which gives the Duffing nonlinearity coefficient of mode-1,

$$Y_1 = \frac{3\alpha_1}{8\omega_1^2} - \frac{\beta_1^2}{2\omega_1^4}, \tag{14}$$

and the mode coupling coefficient between mode-1 and mode-2,

$$\lambda_{1,2} = \frac{EW_{1,1}W_{2,2}}{8\rho L^4 \omega_1^2} - \frac{\beta_1 x_T EW_{2,2}W_{0,1}}{4\rho L^4 \omega_1^4}. \tag{15}$$

When mode-1 is operated under a low excitation, the frequency shift contributed by the $a_1$ is negligible, thus Eq. (13) can be further simplified as

$$\omega_1' - \omega_1 = \Delta\omega = \omega_1 \lambda_{1,2} a_2^2, \tag{16}$$

showing the $\lambda_{1,2}$ enables the oscillation of mode-2 to convert into frequency shift of mode-1. For a certain MEMS resonator, Eq. (12) can be simplified as,

$$\lambda_{1,2} = \frac{A}{\omega_1^2} - \frac{x_T^2}{\omega_1^4} B \tag{17}$$

where A and B are constants determined by the given material and geometry parameters (i.e., $E$, $\rho$ and $L$), leaving on only $x_T$ and $\omega_1$ which are both significantly affected by the thermal strain ($\varepsilon_{th}$) applied to the beam. Thus, $\lambda_{1,2}$ can be significantly tuned by the input heat to the MEMS beam.

It is generally considered that the dispersive mode couplings are mutual, i.e., mode-2 affects mode-1, and mode-1 also affects mode-2 in the same manner. To verify this, we have derived $\lambda_{2,1}$ by switching the roles of mode-1 and mode-2 (i.e., mode-1 for the pump mode and mode-2 for the probe mode). The duffing motion equation then becomes,

$$\ddot{x}_2 + (\omega_2^2 + K_1 x_1^2 + K_2 x_1)x_2 + \alpha_2 x_2^3 = 0, \tag{18}$$

where, $\omega_2^2 = \frac{EI}{\rho SL^4}\psi_2^4 + \frac{T_o W_{2,2}}{\rho SL^2}$, $K_1 = \frac{EW_{1,1}W_{2,2}}{2\rho L^4}$, $K_2 = \frac{x_T EW_{2,2}W_{0,1}}{\rho L^4}$, and $\alpha_2 = \frac{E}{2\rho L^4}W_{2,2}^2$. With the same process as from Eq. (5), we can then obtain $\lambda_{2,1}$ as (see more details in Supplementary Note 2 [37]),

$$\lambda_{2,1} = \frac{EW_{1,1}W_{2,2}}{8\rho L^4 \omega_2^2} - \frac{\beta_1 x_T EW_{2,2}W_{0,1}}{4\rho L^4 \omega_1^2 \omega_2^2}. \tag{19}$$

As seen, $\lambda_{2,1}$ has a very similar form with $\lambda_{1,2}$ and can be modulated by the thermally induced buckling ($x_T$). We further normalize the obtained $\lambda_{1,2}$ and $\lambda_{2,1}$ by their values without any buckling ($x_T = 0$) to obtain the normalized mode coupling coefficients, as

$$\lambda_{1,2}^{N} = \lambda_{2,1}^{N} = 1 - \frac{2\beta_1 x_T W_{0,1}}{W_{1,1} \omega_1^2}. \tag{20}$$

The above Eq. (20) indicates that the buckling effect has the equivalent tuning efficacy for both $\lambda_{1,2}$ and $\lambda_{2,1}$.

## Numerical results

By substituting the geometry and material parameters of the MEMS beam to Eq. (17), we have calculated $\lambda_{1,2}$ at various heating powers, which is plotted as the black curve in Fig. 3(d). As seen, the calculated result shows nice agreement with experimental result. The tuning of $\lambda_{1,2}$ by thermal effect shown in Fig. 3(d) can be understood as follows. When the $\varepsilon_{th}$ increases with $P$ but is much smaller than the Euler's buckling critical strain, $\varepsilon_{cr}$, of the beam ($P< 0.9$ mW), $\omega_1$ decreases while $x_T$ is almost stable, thereby, the $\lambda_{1,2}$ experiences a slight rise. With further increased $P$ approaching the critical buckling point, $\omega_1$ levels off (see Fig. 1(d)) and $x_T$ increases quickly owing to the buckling of the beam [30], giving the sharp reduction in $\lambda_{1,2}$ from the positive to the negative values. However, when $\varepsilon_{th}$ goes beyond $\varepsilon_{cr}$, although $x_T$ still increases with $P$, $\omega_1$ also increases with $P$, giving an almost stable $\lambda_{1,2}$ in the post-buckling regime ($P > 1.8$mW).

Furthermore, we have performed numerical analysis under various initial conditions for achieving strong mode coupling coefficient. Figure 5 (b) plots the theoretically calculated $\lambda_{1,2}$ as a function of normalized strain($\varepsilon_{th}/\varepsilon_{cr}$), at various normalized initial center deflections ($x_0/h$, $h$ is thickness of the beam). As seen, for various $x_0$, the $\lambda_{1,2}$ can be significantly tuned from positive to negative as the $\varepsilon_{th}$ increases, which is constant with the experimental results. Furthermore, the results show that the change of the $\lambda_{1,2}$ highly relies on $x_0$. It is of great interest that, the $\lambda_{1,2}$ can be tuned more significantly with a smaller $x_0$. This is generally owing to the fact that $\omega_1$ and $x_T$ tend to change more dramatically at the critical buckling point for smaller $x_0$ [30]. Figure 5 (c) shows the maximum in the $\lambda_{1,2}$ ($|\lambda_{1,2}^{\max}|$) that can be achieved using thermal tuning as a function of $x_0/h$. When $x_0/h <0.01$, $|\lambda_{1,2}^{\max}|$ increases to ~$2.3\times10^{14}$ m$^{-2}$, which is over 100 times of the measured mode coupling coefficient (~$2.1\times10^{12}$ m$^{-2}$ at $P = 0.9$ mW), enabling the potential in application that needs ultra-strong coupling of vibrational modes.

## Discussion and Conclusion

In this study, we experimentally and theoretically demonstrated the thermal tuning of the mode coupling coefficient in doubly clamped MEMS beam resonators. The tunability of the mode coupling coefficient is promising for applications that require either ultra-strong or weak coupling between vibrational modes in MEMS resonators. One example is to increase the detectable capability of higher-order vibration modes in MEMS resonators [20,23]. With coupling, the vibration of higher-order modes can be probed with the resonance frequency shift of lower-frequency mode, tuning the mode coupling coefficient is, therefore, an effective approach to improve the readout strength of high-frequency modes. This is in particularly important with the scaled NEMS resonators, of which the high-frequency modes commonly have very high resonance frequencies that are difficult to be detected with conventional methods. Another potential application is to achieve multimode sensing using two vibrational modes without cross-talk. Multimode sensors employing two or more vibrational modes enable simultaneous detection of different physical stimulus [38,39]. When frequency shift is used as the detection scheme, the resonance frequencies of each vibration mode should be kept independent of each other, which can be achieved by modulating the coupling coefficient to 0. Furthermore, other than thermal strain, the introduction of the lattice mismatch strain may also an alternative method for tuning the mode coupling coefficient [40], where the compressive strain is preloaded in the beam, thus eliminating the need for an additional heating system.

The theoretical model we established reveals the physical origin of the mode coupling coefficient in MEMS beam resonators. Unlike internal resonance, mode coupling in this study is more generally present in an oscillating MEMS beam resonator and shares the same physical origin with the nonlinearity of two coupled modes, enabling the tuning of mode coupling strength by thermally inducing the buckling of the beam. In terms of design strategies for a doubly-clamped beam structure, a smaller initial center deflection is preferable for achieving a large mode coupling coefficient using strain tuning. On the other hand, for achieving a smaller mode coupling coefficient, the MEMS beam should be operated near its critical buckling point.

Furthermore, from a more essential perspective, both nonlinearity and mode coupling coefficient originate from the additional tension caused by the extension of the beam in vibration, thereby, they are highly correlated with each other. We have derived the analytical relations between the mode coupling coefficient and nonlinearities of the two modes, as

$$\lambda_{1,2} = \frac{\sqrt{\alpha_1 \times \alpha_2}}{4\omega_1^2} - \frac{\beta_1^2 W_{2,2}}{6\omega_1^4 W_{1,1}}, \tag{21}$$

$$\lambda_{2,1} = \frac{\sqrt{\alpha_1 \times \alpha_2}}{4\omega_2^2} - \frac{\beta_1^2 W_{2,2}}{6\omega_1^2 \omega_2^2 W_{1,1}}, \tag{22}$$

where, $\alpha_1$, $\alpha_2$ are the hardening (cubic) nonlinearity coefficients of mode-1 and mode-2, respectively. $\beta_1$ is the softening (quadratic) nonlinearity coefficient of mode-1, which is in proportional to the $x_T$. From Eq. (21) and (22), the $\lambda_{1,2}$ and $\lambda_{2,1}$ can be described as an interaction between the nonlinearities of the two coupled modes. In our previous publication [30], we have demonstrated that the thermal strain can efficiently enhance the $\beta_1$ by increasing $x_T$, thus the tuning in $\lambda_{1,2}$ and $\lambda_{2,1}$ can also be understood by the enhanced $\beta_1$. In addition, Eq. (21) and (22) also provide a method to calculate $\lambda_{1,2}$ and $\lambda_{2,1}$ by simply measuring the Duffing nonlinearities of the two coupled modes.

Eq. (20) indicates that $\lambda_{2,1}$ keeps a synchronous change with $\lambda_{1,2}$ during the tuning, implying a significant transition from positive to negative. To verify this, we have performed the experimental measurement similar to the results shown in Fig. 2. Specifically, we have driven mode-2 in the self-oscillation mode, and excited mode-1 to simultaneously measure the amplitude of mode-1 and frequency of mode-2 at various heating powers. The measurement results are presented in the Supplementary Note 2 [37], from which we can obtain the experimental $\lambda_{2,1}$ as shown by the red dots of Fig. 3(d). As seen, the $\lambda_{2,1}$ has been significantly modulated from positive to negative as heating power increases, aligning well with the tunability of $\lambda_{1,2}$. Furthermore, we have plotted the theoretical $\lambda_{2,1}$ as the red curve in Fig.3 (d), which has a reasonable agreement with the experimental $\lambda_{2,1}$ and provides a clearer trend in the variation of $\lambda_{2,1}$.

It is worth noting that the present theoretical model focuses solely on dispersive mode coupling, whereas the coherent energy transfers between modes at integer frequency ratios are neglected. Specifically, when $f_1/f_2$=1:3, a notable drop in the measured $\lambda_{1,2}$ is observed; however, this feature cannot be explained by the current model. The intrinsic relationship between internal mode resonance and dispersive mode coupling calls for further investigation in future studies.

In conclusion, we have studied the dispersive mode coupling between two flexural vibrational modes of an asymmetric doubly-clamped MEMS beam resonator. We measured the coupling coefficients at various heating powers by measuring the resonance frequency of the probe mode as a function of the oscillation amplitude of the pump mode. The experiment result demonstrates that the coupling coefficients can be significantly tuned by thermally-induced buckling effect, which is promising for controlling the mode coupling strength in MEMS resonators to realize advanced sensing devices. A theoretical model is developed to quantitatively describe the mode coupling coefficient, and shows nice agreement with the experiment result. The proposed model offers insights into fundamental principles of mode coupling in MEMS beam resonators, contributing to understanding mode coupling and the utilization of mode coupling in practical applications.

## Acknowledgments

We thank Dr. Isao Morohashi for the support in the device fabrication process. This work has been partly supported by the A-STEP program of JST, MEXT Grant-in-Aid for Scientific Research on Innovative Areas "Science of hybrid quantum systems" (15H05868), and KAKENHI from JSPS (21K04151, 24K00937). Grants from The Mitsubishi Foundation, The Murata Science Foundation, and The Futaba Foundation are also gratefully acknowledged.

## Author contributions

Ya Zhang conceived the idea of the work. Chao Li designed and fabricated the device, and performed the experiments and data analysis. Qian Liu and Kohei Uchida supported the fabrication and measurement. The theoretical works were done by Chao Li. Chao Li and Ya Zhang wrote the manuscript. All the authors contributed to the discussions in this work.

## Competing interests

The authors declare no competing interests.

**Figures and captions**

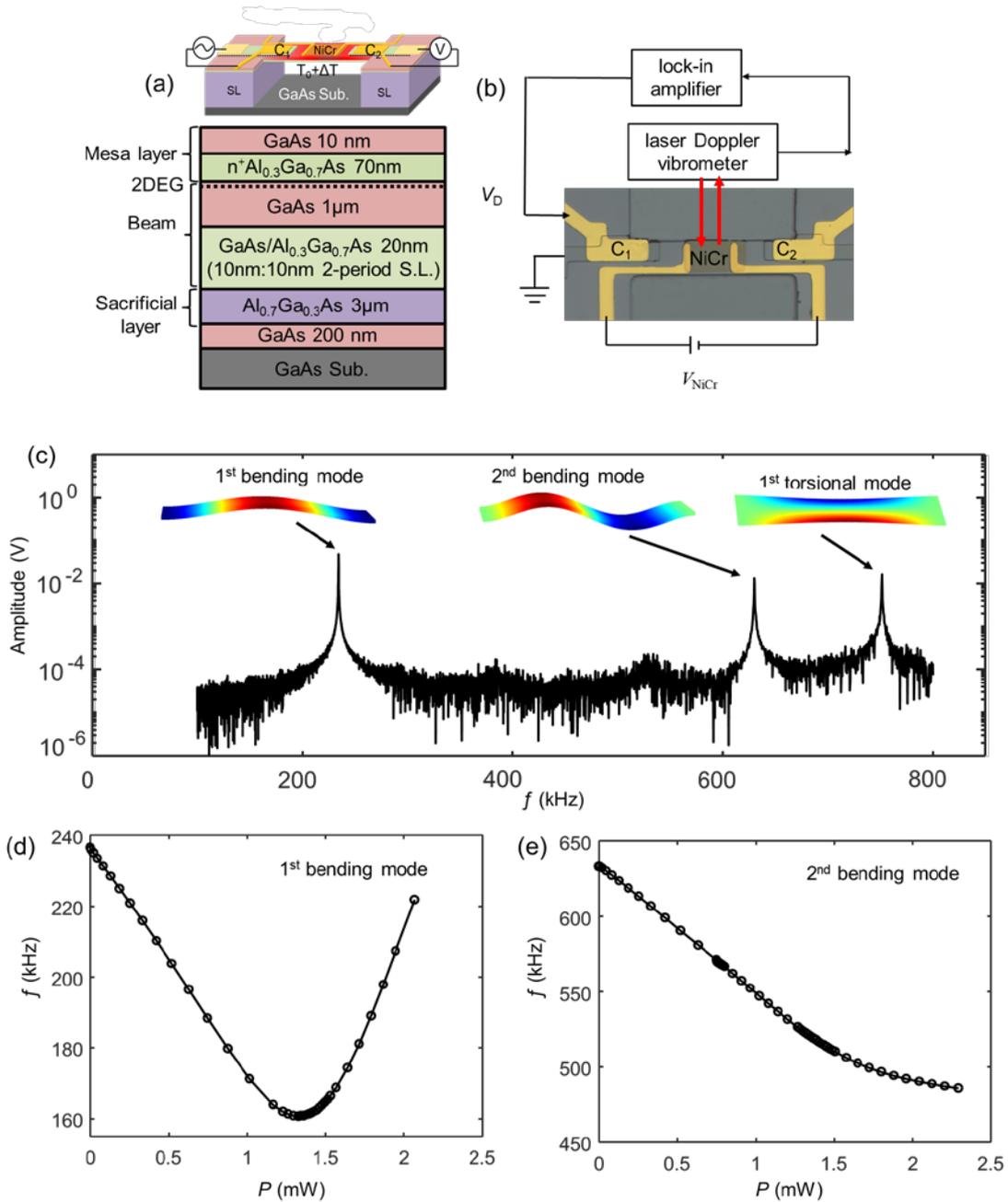

**Figure 1** (a) The wafer structure used for fabricating the GaAs MEMS beam resonators. The upper inset shows the beam structure. (b) A microscope image of fabricated MEMS beam resonator. An ac voltage ($V_D$) is applied to one of the piezoelectric capacitors to drive the resonator and the induced mechanical oscillation is measured by a laser Doppler vibrometer and a lock-in amplifier with a built-in PLL. A dc voltage ($V_{th}$) is applied to the NiCr film to generate heat in the beam. (c) Measured spectrum for the first three modes. The upper insets show the mode shapes of the first three modes, which are obtained by FEM simulation. (d-e) The measured resonance frequencies of the 1$^{st}$ bending mode (d) and the 2$^{nd}$ bending mode (e) as a function of heating power, $P$. The 1$^{st}$ bending mode achieves its critical buckling at $P$=1.33 mW (160 kHz).

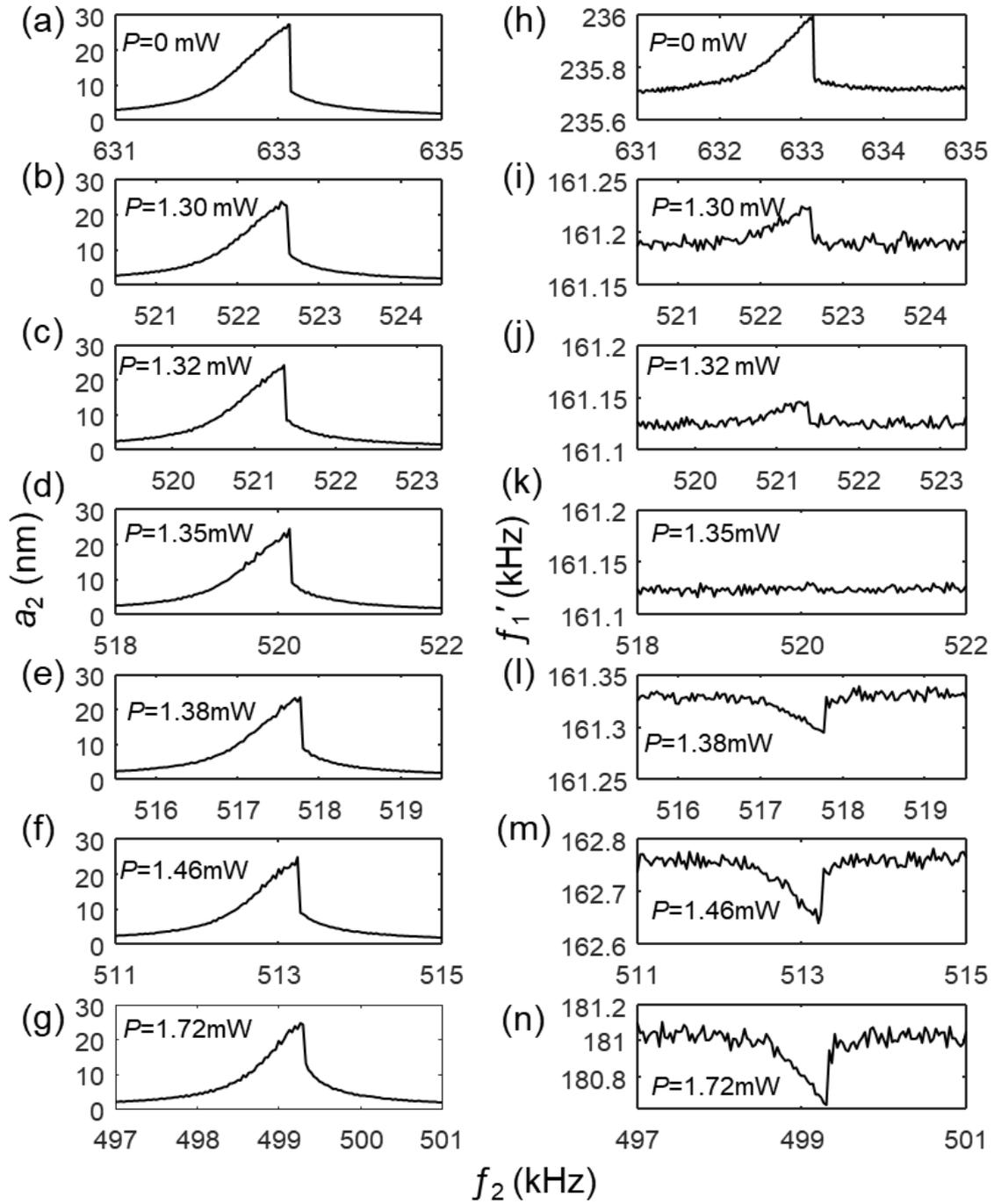

**Figure 2** Coupling measurements between the pump and probe modes using PLL mode, at various heating powers. The x-axis plots the driving frequency (forward sweep) of the mode-2 ($f_2$). (a-g) The oscillation spectra of mode-2, the y-axis are plotted as the oscillation amplitude of the mode-2 ($a_2$). Note that, in the measurement, the laser spot should be focused at 1/4 of the beam length to achieve large oscillation amplitude of mode-2. (h-n) The corresponding resonance frequency of the mode-1 ($f_1'$) with the excitation of the mode-2.

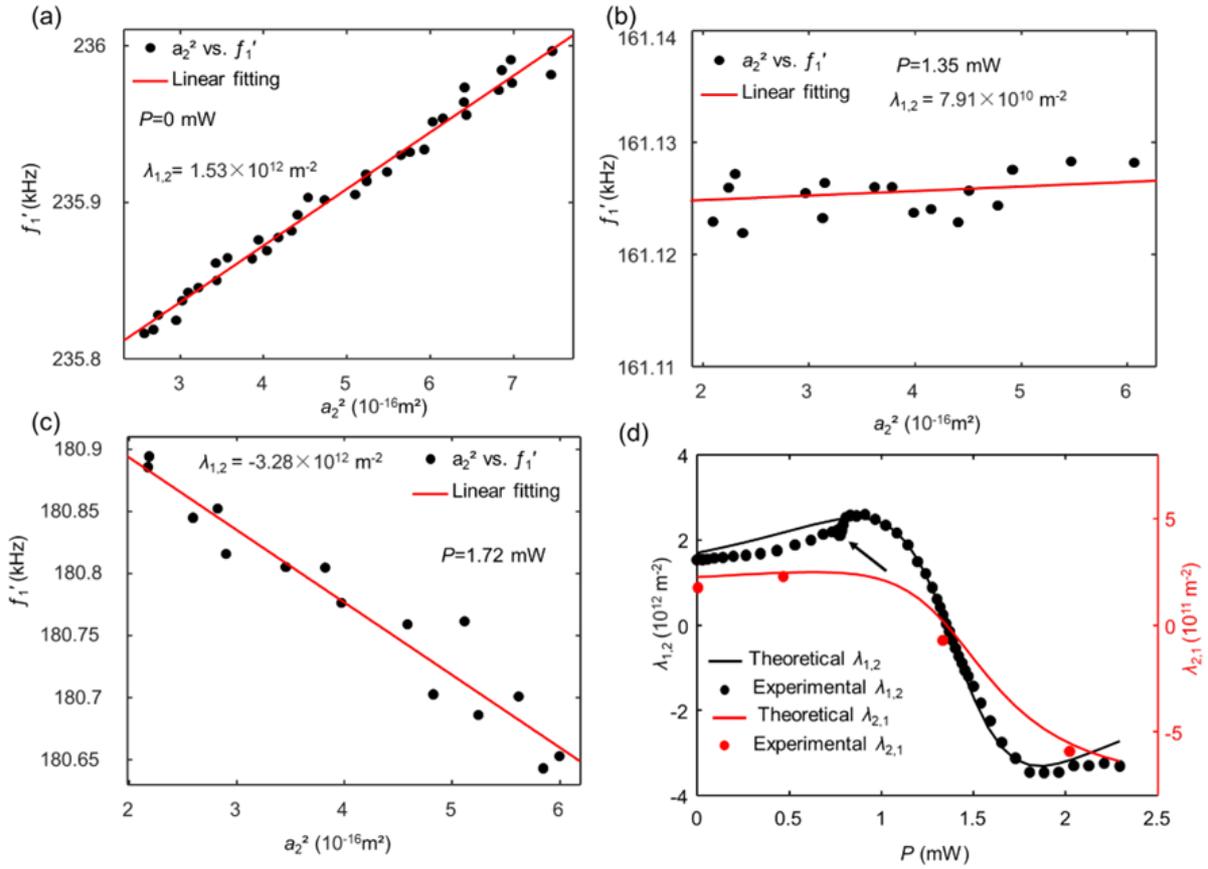

**Figure 3** (a-c) The measured $f_1'$ as a function of the squared amplitude of the pump mode ($a_2^2$) for the cases of $P$=0, 1.35 and 1.72 mW, respectively. (d) The left y-axis plots the experimental mode coupling coefficient, $\lambda_{1,2}$, (black dots) and theoretical $\lambda_{1,2}$ (black line) as a function of heating power. The right y-axis plots the experimental $\lambda_{2,1}$ (red dots) and theoretical $\lambda_{2,1}$ (red line). The black arrow indicates a sudden drop in the measured $\lambda_{1,2}$.

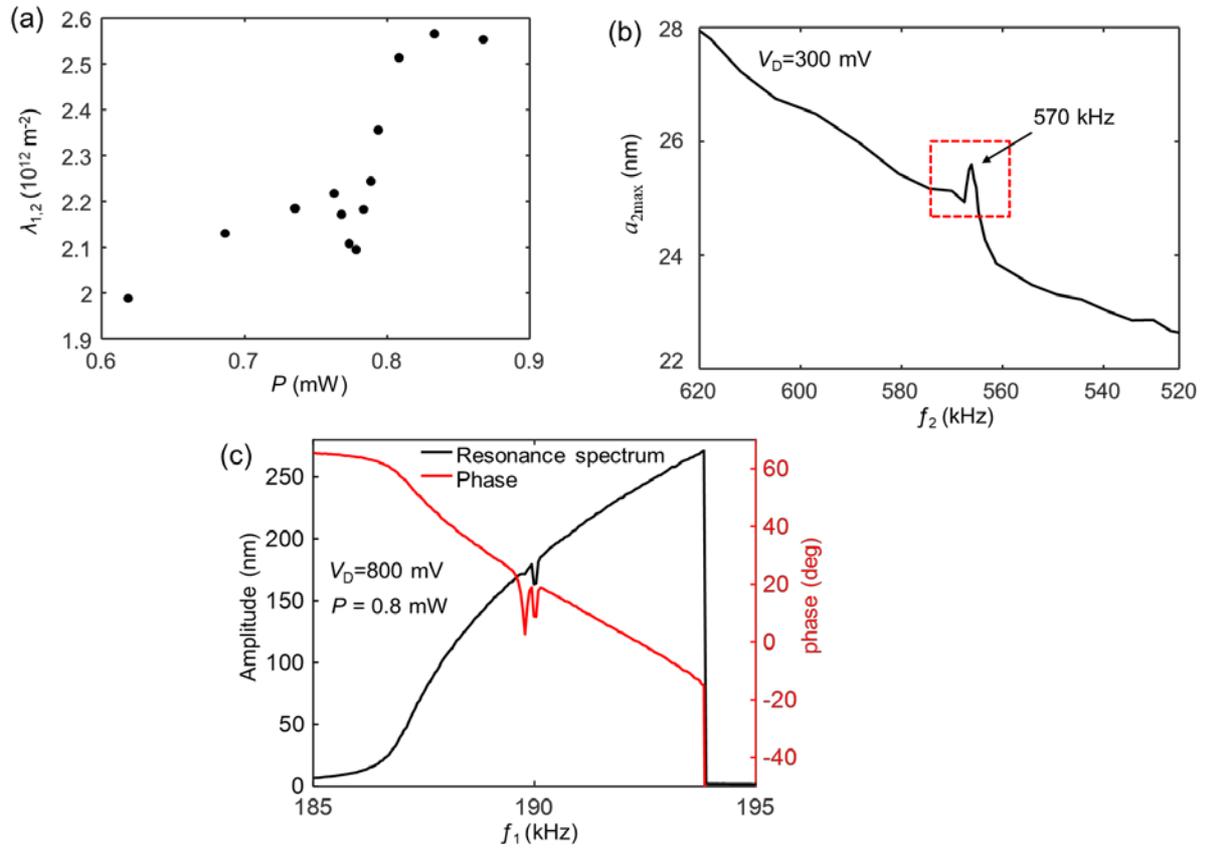

**Figure 4** (a) A blow-up of the measured $\lambda_{1,2}$ at the heating power range of 0.6-0.9 mW. (b) The measured peak oscillation amplitude of mode-2 ($a_{2\text{max}}$) when its resonance frequency modulated from 620 kHz to 520 kHz by heating power. The sudden jump marked by the red dashed rectangle indicates the amplitude enhancement due to 1:3 internal resonance. (c) The resonance spectrum and phase of mode-1 near 190 kHz, with a driving voltage of $V_D$=800 mV. Two clear drops in both spectrum and phase plots indicate the presence of 1:3 internal resonance.

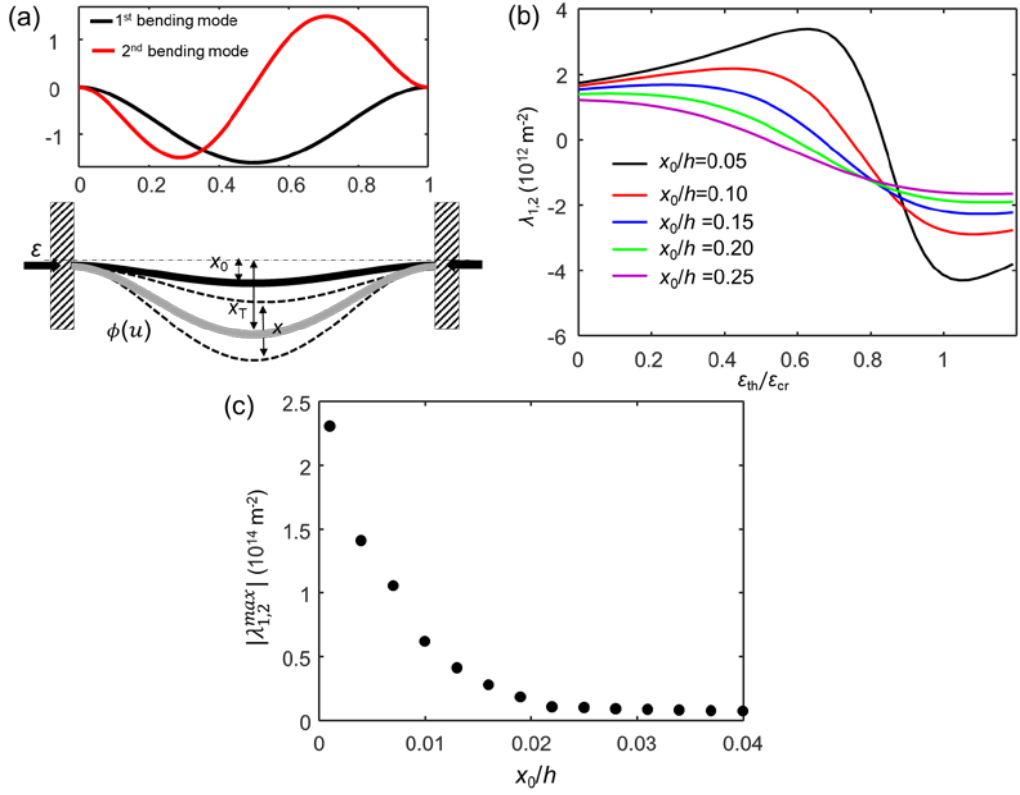

**Figure 5** (a) Schematic diagram of a doubly-clamped MEMS beam with an initial center deflection, $x_0$, which shows an asymmetric structure. When the center deflection increases from $x_0$ to $x_T$, the MEMS beam has a new equilibrium position for the oscillation. The inset shows the 1-dimensional mode shapes of the 1$^{st}$ bending and 2$^{nd}$ bending modes. (b) The theoretically calculated mode coupling coefficient, $\lambda_{1,2}$, as a function of the compressive strain ($\varepsilon_{th}/\varepsilon_{cr}$) at various initial center deflections ($x_0/h$= 0.05, 0.10, 0.15, 0.20, 0.25). $\varepsilon_{th}$ is normalized by the $\varepsilon_{cr}$. (c) The achievable maximum in the mode coupling coefficient ($|\lambda_{1,2}^{max}|$) as a function of $x_0/h$.

24